\documentclass[lettersize,journal]{IEEEtran}
\usepackage{amsmath,amsfonts}
\usepackage{algorithmic}
\usepackage{algorithm}
\usepackage{array}
\usepackage[caption=false,font=normalsize,labelfont=sf,textfont=sf]{subfig}
\usepackage{textcomp}
\usepackage{stfloats}
\usepackage{url}
\usepackage{verbatim}
\usepackage{graphicx}
\usepackage{cite}
\usepackage{comment}
\newtheorem{theorem}{Theorem}

\newtheorem{lemma}{Lemma}

\usepackage{mathtools}
\usepackage{xcolor}
\usepackage{multirow}
\DeclareMathOperator*{\argmax}{argmax}
\DeclareMathOperator*{\argmin}{argmin}
\renewcommand{\baselinestretch}{0.981}

\begin{document}

\title{Relay Selection in Wireless Networks as Restless Bandits}

\author{Mandar R. Nalavade, Ravindra S. Tomar, Gaurav S. Kasbekar,~\IEEEmembership{Member,~IEEE} 
\thanks{The authors are with the Department of Electrical Engineering, Indian Institute of Technology (IIT) Bombay, Mumbai, 400076, India. Their email addresses are 22d0531@iitb.ac.in, 30005388@iitb.ac.in, and gskasbekar@ee.iitb.ac.in. The work of all three authors has been supported in part by the project with code RD/0121-MEITY01-001.}}



\maketitle

\begin{abstract}
We consider a wireless network in which a source node needs to transmit a large file to a destination node. The direct wireless link between the source and the destination is
assumed to be blocked. Multiple candidate relays are available to forward packets from the source to the destination. A holding cost is incurred for each packet stored at every relay in each time slot. The objective is to design a policy for selecting a relay in each time slot to which the source attempts to send a packet, so as to minimize the expected long-run time-averaged total packet holding cost at the relays. This problem is an instance of the
restless multi-armed bandit (RMAB) problem, which is provably hard to solve. We prove that this relay selection problem is Whittle-indexable, and propose a method to compute the Whittle index of each relay in every time slot. In each time slot, our relay selection policy transmits a packet to the relay with the smallest Whittle index. Using simulations, we show that the proposed policy outperforms the relay selection policies proposed in prior work in terms of average cost, delay, as well as throughput.
\end{abstract}


\section{Introduction}

\IEEEPARstart{W}{ireless}channels are vulnerable to a number of impediments, such as path loss, shadow fading, multipath fading, blockages, etc., which limit the performance of wireless networks. An effective and practical technique to overcome such impediments is to use \emph{relays} to forward data packets from a source to a destination \cite{laneman2004cooperative}. Often, multiple nodes are available to act as a relay from a given source to a given destination; in such cases, the relay selection problem, i.e., the problem of deciding which of the available nodes will act as a relay in each time slot, needs to be solved to optimize the network performance \cite{nam2008relay}.  

A number of relay selection schemes have been proposed in prior work \cite{kim2023joint, ju2023deep, ona2020relay, sheng2014energy, zhang2020cooperative, li2023joint, gao2023optimal} for different environments and to achieve various objectives. Deep reinforcement learning (DRL)-based relay selection policies were proposed in  \cite{kim2023joint}, \cite{ju2023deep} for millimeter wave (mmWave) networks, whereas a machine learning algorithm, viz., the multi-user multi-armed bandit algorithm, was proposed in \cite{zhang2020cooperative} for relay selection in wireless sensor networks. An optimal, but exponential-time, Branch-and-Bound (BB)-based relay selection algorithm was proposed in \cite{ona2020relay}, and a power allocation method for decode-and-forward cooperative communication was proposed in \cite{sheng2014energy} for relay selection in wireless multimedia networks. In \cite{li2023joint}, two methods for relay selection and transmission scheduling in mmWave communication were proposed: random relay selection with concurrent scheduling and relay selection with dynamic scheduling. In \cite{gao2023optimal}, a quantum black hole algorithm was proposed to solve the relay selection problem. A Whittle indexability for a class of RMAB processes in a dynamic multichannel access context in \cite{kequin2010indexability} was established and proposed a Whittle index-based policy. 
In \cite{tripathi2019whittle}, a scheduling problem to minimize a function of the age of information was formulated as RMAB problem and solved using the Whittle index-based policy. In \cite{wang2019whittle}, a heuristic policy based on the Whittle index for the RMAB problem of dynamic channel allocation for remote state estimation of multiagent systems was proposed.

We consider a wireless network in which a source node (mobile device) needs to transmit a large file, modeled as an infinite queue of packets, to a destination node, which is a base station (BS). The direct wireless link between the source and the destination is assumed to be blocked. Multiple candidate relays are available to forward packets from the source to the destination. Time is divided into slots of equal duration, with each slot being further subdivided into two mini-slots. The channel quality between the source and each relay, as well as between each relay and the destination, in a time slot is modeled as a random variable. A holding cost is incurred for each packet stored at every relay in each time slot. In the first mini-slot of each slot, a packet is sent from the mobile device to one relay, which is selected from the set of all relays. In the second mini-slot of each slot, a packet is sent from every relay with a non-empty queue to the BS. The objective is to design a relay selection policy that minimizes the expected long-run time-averaged total packet holding cost at the relays. Such a policy minimizes the average packet delay at the relays. 
This problem is an instance of the \emph{restless multi-armed bandit (RMAB) problem}, which is provably hard to solve \cite{papadimitriou1994complexity}.  
We prove that this relay selection problem is Whittle indexable \cite{whittle1988restless} and propose a method to compute the Whittle index of each relay in every time slot.
In each time slot, our relay selection policy transmits a packet to the relay with the smallest Whittle index. Using simulations, we demonstrate that the proposed policy outperforms the relay selection policies proposed in prior work in terms of the average cost, delay, and throughput. 

The closest to this paper is \cite{kaza2018restless}, where an opportunistic scheduling problem in communication systems and relay selection is formulated as partially observable RMAB with cumulative feedback and it is solved using the Whittle index policy. However, our paper is significantly different from \cite{kaza2018restless} in the following aspects: (i) In the system model of our paper, a time slot is further subdivided into two mini-slots, whereas no further subdivision of a slot is considered in \cite{kaza2018restless}. (ii) The channel is assumed to be a Bernoulli random variable in our paper, whereas the channel is modeled as a Gilbert–Elliot model in \cite{kaza2018restless}. (iii) The states in our paper are fully observable, but the states are partially observable in \cite{kaza2018restless}. (iv) In our paper, the problem of relay selection is formulated as an RMAB problem with per-slot constraints with the objective of minimizing long-run expected average holding cost. In \cite{kaza2018restless}, the relay selection problem is formulated as a partially observable RMAB problem with cumulative feedback and the objective of maximizing the discounted cumulative reward (throughput). (v) The work in \cite{kaza2018restless} generalizes RMAB by introducing belief updates and providing closed-form Whittle indices for special cases, our contribution is to apply Whittle index theory directly to the relay selection problem. Our key contribution lies in proving the Whittle indexability of the relay selection problem and proposing a computational method for index evaluation. (vi) Furthermore, our evaluation is tailored to relay networks and demonstrates that the proposed Whittle index-based policy consistently outperforms existing relay selection strategies such as random, load-based, Max-Min Relay Selection (MMRS), and Max-Link Relay Selection (MLRS) in terms of average cost, delay, and throughput, while \cite{kaza2018restless} is primarily focusing on throughput maximization and comparison with generic policies such as myopic, random, and round-robin. (vii) Computational complexity is specified in our paper, whereas in \cite{kaza2018restless}, it is absent. To the best of our knowledge, except \cite{kaza2018restless}, the powerful technique based on the \emph{Whittle index} \cite{whittle1988restless} has not been used in prior work to solve the relay selection problem in wireless networks. 

The key contributions of this paper are given below. (i) We modeled the relay selection problem in a wireless network as a RMAB problem. 
Note that, in general, it is difficult to prove the Whittle indexability of a RMAB problem. One of the key contributions of this letter is that we prove the problem of relay selection in wireless networks to be Whittle indexable. Except \cite{kaza2018restless}, the Whittle index technique has not been used for relay selection in prior works. (ii) After the start of network operation, according to the proposed Whittle index-based relay selection policy, a relay with the lowest Whittle index needs to be selected in each time slot. Hence, this strategy is computationally more effective. A detailed computational complexity analysis is provided. (iii) Furthermore, our evaluation is tailored to relay networks and demonstrates that the proposed Whittle index-based policy consistently outperforms existing relay selection strategies in terms of average cost, delay, and throughput. 

The rest of this letter is structured as follows. In Section \ref{system_model}, we describe the system model and problem formulation. In Section \ref{whittle_index_relay}, we prove the Whittle indexability of the relay selection problem. In Section \ref{whittle_computation}, we propose a method to compute Whittle indices, and analyze the computational complexity of the proposed method. We present simulation results in Section \ref{Section_simulation}. Finally, we provide conclusions and directions for future research in Section \ref{Section_conclusion}.

\section{System Model And Problem Formulation}\label{system_model}
\begin{figure} [!t]
\centerline{\includegraphics[width=\linewidth]{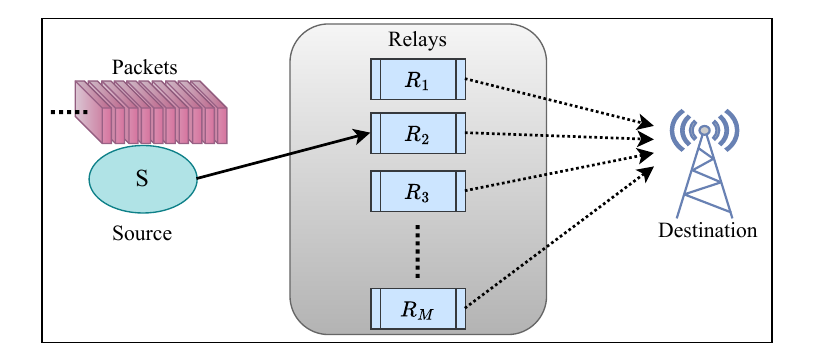}}
\caption{The figure illustrates the system model.}
\label{fig:sys_model}
\end{figure}

Consider a source, $S$, which is a mobile device that wants to send a large file to a destination $D$, which is a BS. The direct wireless link from $S$ to $D$ is blocked. Hence, $S$ uses a relay to communicate with $D$.  There are $M$ candidate relays, $R_1, \ldots, R_M$ (see Fig. \ref{fig:sys_model}). We model the large file that needs to be transferred from $S$ to $D$ by assuming that there is an infinite queue of packets at $S$ that need to be sent to $D$. Time is divided into slots $n \in \{0, 1, 2, \ldots\}$. 
Each slot is divided into two mini-slots. In the first mini-slot of each slot, a packet is sent from $S$ to one of the relays. In the second mini-slot of each slot, a packet is sent from every relay with a non-empty queue to $D$.

In the first mini-slot of each slot, $S$ needs to choose a relay to which to attempt to send a packet. The channel from $S$ to $R_i$ (respectively, from $R_i$ to $D$) in the first (respectively, second) mini-slot of a time slot is a Bernoulli random variable with success probability $f_i$ (respectively, $l_i$). That is, if a packet is sent from $S$ to $R_i$ (respectively, $R_i$ to $D$), then it is successfully received at $R_i$ (respectively, $D$) with probability (w.p.) $f_i$ (respectively, $l_i$). Let $X_n^i$ denote the queue length at relay $R_i$ at the beginning of slot $n$. The queue length at relay $R_i$ evolves according to the following equation: 
\begin{equation} \label{state_update_equation}
     X_{n+1}^i = \left (X_n^i + \mu_n^i A_n^i - W_n^i\right )^+,    
\end{equation} 
where $y^+ \coloneq \max (y,0)$, $\mu_n^i$ is a binary decision variable, which indicates whether a packet is sent from $S$ to relay $R_i$  in time slot $n$ or not and is given by:
\[
\mu_n^i = \left\{ 
\begin{array}{ll}
  1,   & \mbox{if a packet is sent from } S \mbox{ to relay } R_i \\
        & \mbox{in the first mini-slot of slot } n,  \\
  0,   &  \mbox{else,} 
\end{array} \right.
\]
$A_n^i$ is a Bernoulli random variable, which is given by:
\begin{equation*} 
A_n^i=\begin{cases}
        1,  & \mbox{if the quality of the channel from } S \mbox{ to } R_i \\
        & \mbox{is good in time slot } n,  \\
        0, & \mbox{else},
       \end{cases}
\end{equation*}
and $W_n^i$ is a Bernoulli random variable, which is given by:
\begin{equation*} 
W_n^i=\begin{cases}
        1,  & \mbox{if the quality of the channel from } R_i \mbox{ to } D \\
        & \mbox{is good in time slot } n,  \\
        0, & \mbox{else}.
       \end{cases}
\end{equation*}

Let $C_i > 0$ be the per-slot packet holding cost at relay $R_i$. In time slot $n$, the overall cost experienced by all the relays in the network is $\sum_{i=1}^{M} C_i X_n^i$. In this letter, our goal is to devise a non-anticipating policy \cite{hordijk1983average},  which selects appropriate controls $\mu_n^i, i \in \{1, \ldots, M\}$, for each time slot $n$, aiming to minimize the long-term expected average cost incurred by the relays. The problem is:
\begin{align} \label{primary_objective}
 \min  \underset {T \uparrow \infty} {\text{ limsup }} E \left[ \frac{1}{T} \sum_{n=0}^{T-1}\sum_{i=1}^{M}C_i X_{n}^{i}\right],\;\; \text{s.t.} \sum_{i=1}^{M} \mu_{n}^{i} = 1, \forall n.
\end{align}
Note that minimizing the average cost in \eqref{primary_objective} causes the average packet delays at the relays to be small. 

The exact per-stage  constraint, $\sum_{i=1}^{M} \mu_{n}^{i} = 1$, makes it provably difficult to solve the above constrained optimization problem \cite{papadimitriou1994complexity}. Hence, we employ a relaxation technique proposed by Whittle \cite{whittle1988restless}, and relax this constraint to the following time-averaged constraint:
\begin{equation} \label{relaxed_constraint}
\underset {T \uparrow \infty} {\text{ limsup }} \frac{1}{T} \sum_{n=0}^{T-1} \sum_{i=1}^{M}E \left [\mu_{n}^{i}\right ] = 1.
\end{equation}
Utilizing the Lagrangian multiplier approach \cite{borkar2002convex}, the original problem in \eqref{primary_objective} is transformed into the following unconstrained optimization problem:
\begin{align} \label{unconstrained_problem}
    \min &\underset {T \uparrow \infty} {\text{ limsup }} \frac{1}{T} \sum_{n=0}^{T-1} \sum_{i=1}^{M}E \left [\mathcal{F}_i(X_{n}^{i}, \mu_{n}^{i})\right ], \notag\\
    \text{where } &\mathcal{F}_i (x,\mu) = C_ix + \lambda \left ( 1 - \mu \right ),
\end{align}
and $\lambda$ is the Lagrangian multiplier. As in the analysis of Whittle \cite{whittle1988restless}, and since the above problem is a cost minimization problem, $\lambda$ can be treated as a tax or penalty. In the context of Whittle indexability, this implies that when a source $S$ does not send a packet to relay $R_i$ (i.e., when $\mu_n^i = 0$), a tax or penalty of $\lambda$ is added to the relay's incurred cost. The minimization problem formulated in \eqref{unconstrained_problem} decomposes into individual control problems for each relay when $\lambda$ is fixed. The control problem corresponding to relay $R_i$ can be modeled as a Markov Decision Process (MDP) with state $X_{n}^{i}$ and action $\mu_{n}^{i}$ and is given by:
\begin{align} \label{final_objective_modified}
    \min &\underset {T \uparrow \infty} {\text{ limsup }} \frac{1}{T} \sum_{n=0}^{T-1}  E \left [  C_i X_n^i + \lambda \left ( 1 - \mu_n^i \right ) \right], \notag \\ &\text{s.t.  } \mu_n^i \in \{0,1\}, \; \forall n.
\end{align}
This MDP framework is said to exhibit Whittle indexability if, for all feasible parameter sets $(C_i, f_i, l_i) \in (0, \infty) \times (0, 1) \times (0, 1)$, and for each relay $R_i$, the collection of states in which a packet is not sent to relay $R_i$ (hereafter termed ``passive states") exhibits a monotonic decrease, transitioning from being the entire state space to becoming the empty set as we increase $\lambda$ from $-\infty$ to $\infty$. For each relay, the Whittle index associated with state $x$ represents the specific $\lambda$ value at which the controller is indifferent between sending and not sending a packet to the relay, i.e., between activity and passivity, respectively. Under our proposed Whittle index-based policy, in the first mini-slot of each slot, the source $S$ transmits a packet to the relay with the lowest Whittle index.

\section{Whittle Indexability}\label{whittle_index_relay}
For the stability of the system at each relay, we assume that $\min (l_1, \ldots, l_M) > \max (f_1,  \ldots, f_M)$. Theorem \ref{TH:stability} below proves the stability of the system at each relay. 

A given control policy for the MDP in \eqref{final_objective_modified} induces a Discrete-Time Markov Chain (DTMC). The state of this induced DTMC is the number of packets at relay $R_i$, and the set of all such states, i.e., the state space, $\mathbb{S}$, is  the collection of all non-negative integers.
\begin{theorem}
\label{TH:stability}
    If $\min (l_1, \ldots, l_M) > \max (f_1, \ldots, f_M)$, then the DTMC induced at each relay $R_i$ is positive recurrent.
\end{theorem}
\begin{IEEEproof}
    Using Proposition 5.3 on p. 21 of \cite{asmussen2003applied}, we can analyze the positive recurrence of the DTMC induced at relay $R_i$. The positive recurrence of the DTMC can be shown by proving the following conditions:
\begin{align*}
    \inf_{x \in \mathbb{S}} \Omega(x) &> -\infty,\\
    \sum_{y \in \mathbb{S}} p_{xy} \Omega(y) &< \infty, \;\forall x \in \mathbb{S}_0, \\
    \Delta \Omega(x) =\sum_{y \in \mathbb{S}} p_{xy} (\Omega(y)-\Omega(x)) &\leq -\delta, \;\forall x \notin \mathbb{S}_0, 
\end{align*}
where $\mathbb{S}_0$ is a finite set such that $\mathbb{S}_0 \subset \mathbb{S}$, $p_{xy}$ is the transition probability from state $x$ to state $y$, $\Omega(x)$ is a Lyapunov function such that $\Omega : \mathbb{S} \rightarrow \mathbb{R}$, and $\delta > 0$. Under the condition given above, i.e., $\min (l_1, \ldots, l_M) > \max (f_1,  \cdots f_M)$, and for the Lyapunov function $\Omega(x) = x$, we can easily prove the above conditions. We omit the details due to space constraints.
\end{IEEEproof}

The results of the paper are applicable to real-world networks with heterogeneous relays, since the relay parameters $f_i, l_i,$ and $C_i$ can differ across nodes. Moreover, the proposed scheme can handle time-varying relays by dividing time into intervals where parameters remain constant, and applying the relay selection policy separately within each interval.

Recall that the problem \eqref{unconstrained_problem} can be decomposed into separate MDPs at different relays for a given value of $\lambda$. 
At the relay $R_i$, let $X_{n}^{i}= x$ denote the current state of the system in slot $n$. We first formulate the Dynamic Programming Equation (DPE) for the individual MDP \eqref{final_objective_modified} at relay $R_i$, and then prove its validity. The DPE for the value function, $V(\cdot)$, of the individual MDP \eqref{final_objective_modified} at relay $R_i$ is given by:
\begin{align} \label{val_fun}
V(x) =& C_i x - \sigma + \min \biggl ( \biggl \{ (1-f_{i})(1-l_{i}) + f_{i}l_{i} \biggl \} V(x) \notag \\
&+(1-f_{i})l_{i}V\left((x-1)^+\right) + f_{i}(1-l_{i})V(x+1); \notag \\
&\lambda + (1-l_{i})V(x) + l_{i}V\left((x-1)^+\right) \biggl ).
\end{align}
In \eqref{val_fun}, $\sigma$ is a constant, whose value we provide later. The first (resp., second) term in the argument of `min' in \eqref{val_fun} corresponds to the activity (resp., passivity) of relay  $R_i$, i.e., $\mu_n^i = 1$ (resp., $\mu_n^i = 0$). Let us prove \eqref{val_fun}. 
The infinite horizon $\gamma$-discounted cost for the controlled MDP associated with relay $R_i$, starting from initial state $x$ under a stationary control policy $\pi$ is given by:
\begin{equation*}
    H^{\gamma}(x, \pi) = E \left [  \sum_{n=0}^{\infty} \gamma^n((1-\mu_n^i)\lambda + C_i X_n^{i})|X_0^{i} = x \right ].
\end{equation*}
Minimizing over all possible stationary control policies $\pi$ yields the value function of the above infinite horizon $\gamma$-discounted problem. It can be expressed as:
\begin{equation*}
    V^{\gamma}(x) = \inf_{\pi} H^{\gamma}(x,\pi).
\end{equation*}
If $q_{.|.}(0)$ and $q_{.|.}(1)$ are the transition probability functions under the controls $\mu^i = 0$ and $\mu^i = 1$, respectively, then the value function of the controlled DTMC induced at relay $R_i$ is given by the following DPE:
\begin{equation*}
    V^{\gamma}(x) =  \min_{\mu^i \in \{0,1\}} \left [ C_ix + (1-\mu^i)\lambda + \gamma\sum_{y}^{}q_{y|x}(\mu^i)V^{\gamma}(y) \right].
\end{equation*}
Let $\bar{V}^{\gamma}(\cdot) = V^{\gamma}(\cdot) - V^{\gamma}(0)$. We can express the above equation as follows:
\begin{equation} \label{beta_val_fun_bar}
\begin{split}
    \bar{V}^{\gamma}(x) =  &\min_{\mu^i \in \{0,1\}} \biggl [ C_ix + (1-\mu^i)\lambda - (1-\gamma)V^{\gamma}(0)  \\
 &+ \gamma\sum_{y}^{}q_{y|x}(\mu^i)\bar{V}^{\gamma}(y) \biggl ].
\end{split}
\end{equation}
The value function of the $\gamma$-discounted problem specified in \eqref{beta_val_fun_bar} is useful in the calculation of the value function $V(\cdot)$ and the constant $\sigma$ in \eqref{val_fun}. These quantities are determined using the following lemma.
\begin{lemma}  \label{lem:lemma1}
$V(\cdot)$ and $\sigma$  in \eqref{val_fun} can be obtained from: $\lim_{\gamma \uparrow 1} \bar{V}^{\gamma}(\cdot) = V(\cdot)$ and $\lim_{\gamma \uparrow 1} (1-\gamma) V^{\gamma}(0) = \sigma$. The constant $\sigma$ in \eqref{val_fun} is unique and equals the optimal long-term expected average cost of the individual MDP \eqref{final_objective_modified} induced at each relay. With the additional constraint $V(0) = 0$, and under an optimal policy, $V(\cdot)$ also maintains uniqueness in states that are positive recurrent. For a given state $x$, the optimal choice of $\mu$ is obtained by finding the argmin of the RHS of (\ref{val_fun}). 
\end{lemma}
\begin{IEEEproof}
    This lemma can be proved using a similar line of reasoning as in the proof of Lemma 4 in \cite{vsbspatta}, with the difference being that we utilize the Lyapunov function introduced in the proof of Theorem \ref{TH:stability}, which differs from the one used in \cite{vsbspatta}. For brevity, the details of the proof are omitted.
\end{IEEEproof}

\begin{theorem}
    The relay selection problem is Whittle indexable.
\end{theorem}
\begin{IEEEproof}
     A user association problem in a dense mmWave network was shown to be Whittle indexable in \cite{singh2022user}. In the system model of \cite{singh2022user}, there are $M$ mmWave BSs (mBSs) in a region. In each time slot, a user arrives w.p. $p$ and a user associated with mBS $i$ departs w.p. $r_i$, where $r_i$ is the service rate from mBS $i$ to each user.
    Every arriving user must be assigned to exactly one of the $M$ mBSs. In \cite{singh2022user}, the objective was to minimize the long-run expected average cost incurred at all the mBSs in the system, and the minimization problem was as in \eqref{primary_objective} in this letter, with the difference that $X_n^{i}$ and $C_{i}$ denote the number of users associated with mBS $i$ at the beginning of slot $n$, and the holding cost  per slot per user incurred at mBS $i$, resp., and:
    \[
\mu_n^i = \left\{ 
\begin{array}{ll}
  1,   & \mbox{if mBS } i \mbox{ admits an arrival in slot  } n,  \\
  0,   &  \mbox{else.} 
\end{array} \right.
\]
This minimization problem was decoupled into separate MDPs for different mBSs following the procedure of Whittle \cite{whittle1988restless} and it was shown in \cite{singh2022user} that the DPE of the MDP of mBS $i$ is given by  (equation (5) in \cite{singh2022user}):
    \begin{align} \label{val_fun_singh}
        V(x) =&  C_ix - \sigma + \min \biggl ( \biggl [ (1-p)(1-r_{i}) + p r_{i} \biggl ] V(x) \notag \\
        &+(1-p)r_{i}V\left((x-1)^+\right) + p(1-r_{i})V(x+1); \notag \\
        &\lambda + (1-r_{i})V(x) + r_{i}V\left((x-1)^+\right) \biggl ).
    \end{align}
    Note that this DPE is equivalent to the DPE \eqref{val_fun} in this letter. Since the proof of Whittle indexability depends only on the DPE, from the proof of Whittle indexability of the user association problem in \cite{singh2022user}, it follows that the relay selection problem in this letter is Whittle indexable. 
\end{IEEEproof}

\section{Whittle Index Computation}\label{whittle_computation}
For a state $x$ of relay $R_i$, the Whittle index $\lambda$ is computed iteratively. Specifically, with $\tau \geq 0$ being the iteration number, the value of $\lambda$ is updated according to the following equation:
\begin{align} \label{update_lambda}
    \lambda_{\tau+1} =& \lambda_{\tau} + \beta \biggl ( \sum_{j}^{}p_{sp}(j|x)V_{\lambda_{\tau}}(j) - \sum_{j}^{}p_{np}(j|x)V_{\lambda_{\tau}}(j) \notag\\
    &- \lambda_{\tau} \biggl ), \; \tau \geq 0,
\end{align}
where $\beta >0$ is a small step size, and $V_{\lambda_{\tau}}(\cdot)$ is the value function, which appears in the DPE \eqref{val_fun}, for the tax $\lambda_{\tau}$. $p_{sp}(\cdot|x)$ and $p_{np}(\cdot|x)$ denote the transition probabilities under the events that the source $S$ sends a packet and does not send a packet to the relay $R_i$, respectively, in the current slot, given the present state $x$. The above iteration \eqref{update_lambda} for $\lambda$ reduces the difference between the two arguments in the minimization function specified in \eqref{val_fun}. So, from \eqref{update_lambda}, we observe that $\lambda$ converges to a value at which the source $S$ obtains equal expected utilities by sending and not sending a packet to the relay $R_i$.

We find $V_{\lambda}$ using a linear system of equations, which can be solved using the present value of $\lambda$ after each iteration of \eqref{update_lambda}. In particular,  at each relay $R_i$, we solve the following system of equations for $V = V_{\lambda_{\tau}}$ and $\sigma = \sigma(\lambda_{\tau})$,  using the value of the parameter $\lambda = \lambda_{\tau}$:

\begin{subequations} \label{sys_linear_eq}
    \begin{align}
    V(y) &=  C_iy - \sigma + \sum_{z}^{} p_{sp}(z|y) V(z), y \leq x, \label{10a}\\
    V(y) &=  C_iy + \lambda - \sigma + \sum_{z}^{} p_{np}(z|y) V(z), y > x, \label{10b} \\
    V(0) &= 0. \label{10c}
    \end{align}
\end{subequations}
Given a state $x$, the Whittle index of the relay $R_i$ is the value of $\lambda$ after the convergence of the iteration \eqref{update_lambda}.
To reduce the computational cost, we execute the iteration \eqref{update_lambda} only for a subset of the states $x$ and then, to find the Whittle indices for the remaining states, we use interpolation. Under the proposed Whittle index-based policy, the relay that has the smallest Whittle index receives the packet sent by the source $S$ in each time slot.

This paragraph focuses on analyzing the complexity of the proposed Whittle index-based user association scheme. For a given state $x$, let $B$ be the maximum number of iterations required for the convergence of \eqref{update_lambda} to the corresponding Whittle index.  Also, let $K_i$ be the capacity of the $i^{th}$ relay, i.e., the maximum queue length at the $i^{th}$ relay at a certain time instant. After each iteration of \eqref{update_lambda}, the system of equations given in \eqref{sys_linear_eq} is solved. This represents the system of $K_i + 1$ linear equations with $K_i+1$ unknowns in $V(y)$, $y \in \{0, 1, \ldots, K_i\}$. Thus, the complexity of solving this system of linear equations for the unknowns $V(\cdot)$ is $\mathcal{O}(K_i^3)$. Hence, for each iteration of \eqref{update_lambda}, the complexity incurred is $\mathcal{O}(K_i^3)$. Therefore, the complexity of the calculation of the Whittle index for an individual relay $i$ and for a given state $x$ is $\mathcal{O}(B \times K_i^3)$. So, the complexity of computing the Whittle indices for an individual relay $i$ and all states in the state space is $\mathcal{O}(B \times K_i^4)$. Finally, the overall complexity of calculating the Whittle indices for all relays and all states of each relay is $\mathcal{O}(B \sum_{i=1}^{M} K_i^4)$. Note that this is a one-time computation performed before the network operation starts. Subsequently, in every time slot $n \in \{0, 1, 2, \ldots\}$, we need to look up the Whittle index for each relay and select the relay with the smallest Whittle index. So, the computational cost in every time slot is $\mathcal{O}(M)$. 
Hence, the proposed scheme is highly scalable and directly applicable to real-time applications without the need for any approximations or heuristic improvements.

\section{Simulations}\label{Section_simulation}
In Section \ref{SSC:other:policies}, we briefly describe various relay selection policies proposed in prior work. We use MATLAB simulations to assess how the proposed Whittle index-based relay selection policy performs relative to those briefly described in the Section \ref{SSC:other:policies}.

\subsection{Other Existing Policies}
\label{SSC:other:policies}

\subsubsection{Random Policy}
In the first mini-slot of each time slot, the source $S$ sends a packet to a relay, which is selected uniformly at random out of  the $M$ relays. 

\subsubsection{Load Based Policy}
In each time slot, the source $S$ sends a packet to the relay with the smallest packet queue length at the start of the time slot, with ties resolved randomly. That is, in the first mini-slot of time slot $n$, the source $S$ sends a packet to  relay $R_i$, where $i = \argmin_{j \in \{1,\ldots,M\}} X_n^j$.

\subsubsection{Max-Min Relay Selection (MMRS) Policy}
Under the MMRS policy \cite{shen2021ber}, in the first mini-slot  of each time slot, the source $S$ transmits a packet to the relay that offers the highest minimum channel quality of the path from $S$ to the destination $D$ via the relay, with ties resolved randomly. That is, the MMRS policy selects the relay $R_i$, where $i = \argmax_{j \in \{1,\ldots,M\}} \min \left \{  f_j, l_j\right\}$.

\subsubsection{Max-Link Relay Selection (MLRS) Policy}
Under the MLRS policy \cite{el2020buffer}, the relay that has the highest product of queue length and quality of the channel from the relay to the destination $D$ is selected, with ties resolved randomly. That is, in the first mini-slot of time slot $n$, $S$ transmits a packet to relay $R_i$, where $i = \argmax_{j \in \{1,\ldots,M\}} X_n^j l_j$.

\subsection{Simulation Setup and Results}
\label{SSC:simulation:setup:results}
\begin{figure}[!t]
\centering
\subfloat[]{\includegraphics[width=0.24\textwidth]{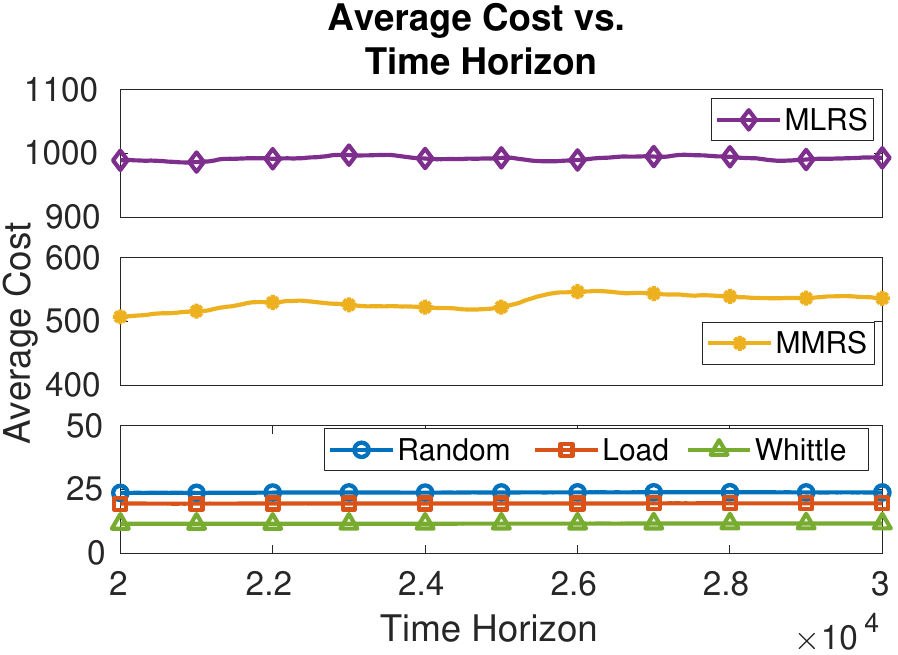}%
\label{fig:2a}}
\hfil
\subfloat[]{\includegraphics[width=0.24\textwidth]{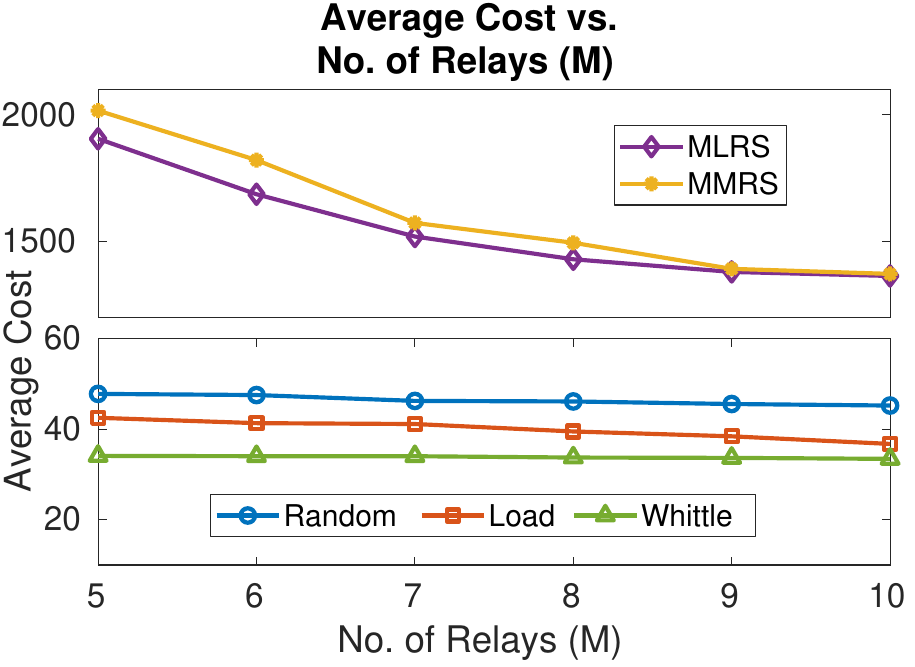}%
\label{fig:2b}}
\caption{The figures show  comparisons of the average costs incurred under the five relay selection policies. The parameter values used for figures are described in Table \ref{tab:parameters_fig_2_to_5}. 
}
\label{fig:2}
\end{figure}

\begin{figure}[!t]
\centering
\subfloat[]{\includegraphics[width=0.24\textwidth]{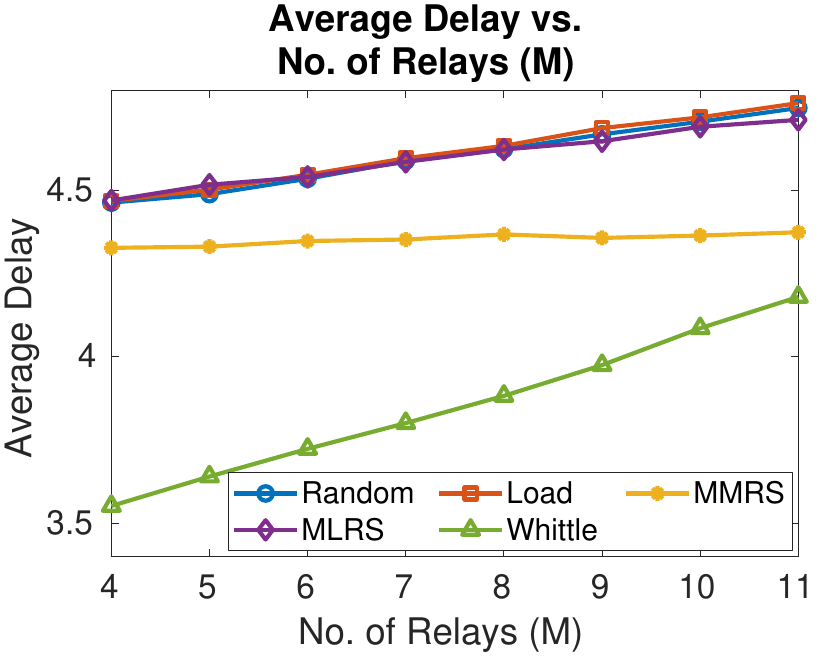}%
\label{fig:3a}}
\hfil
\subfloat[]{\includegraphics[width=0.24\textwidth]{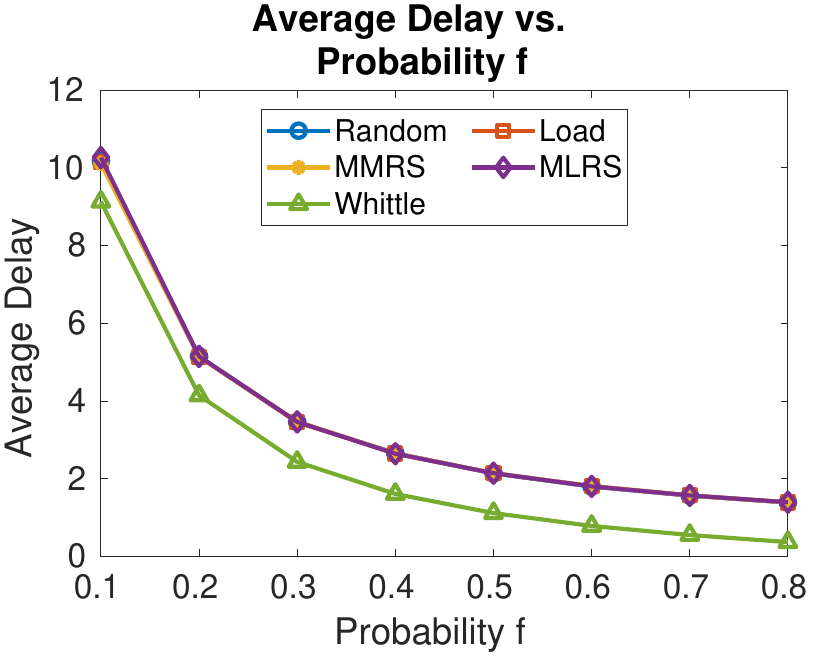}
\label{fig:3b}}
\caption{The figures show comparisons of the average delays incurred under the five relay selection policies. The parameter values used for  figures figures are described in Table \ref{tab:parameters_fig_2_to_5}. 
}
\label{fig:3}
\end{figure}

\begin{figure}[!t]
\centering
\subfloat[]{\includegraphics[width=0.24\textwidth]{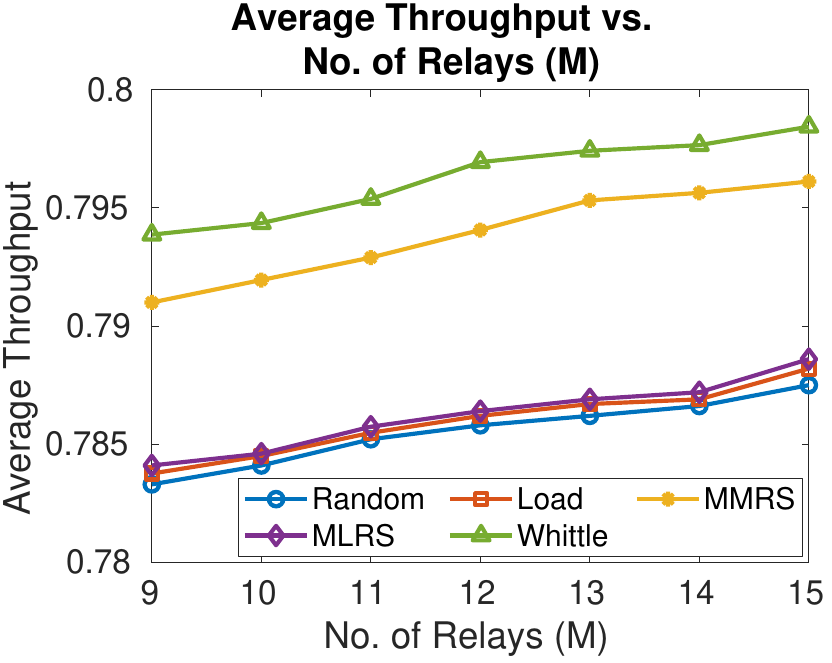}%
\label{fig:4a}}
\hfil
\subfloat[]{\includegraphics[width=0.24\textwidth]{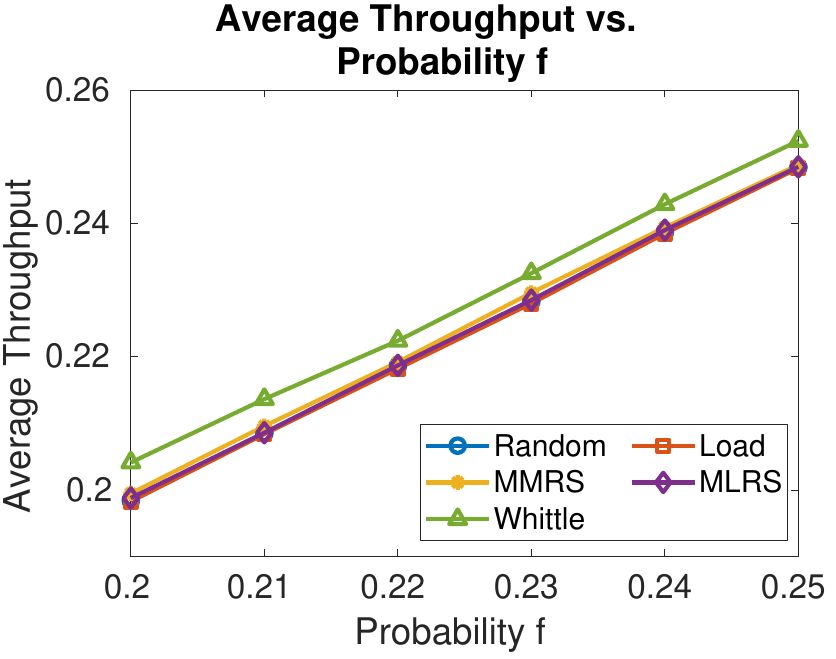}%
\label{fig:4b}}
\caption{The figures show comparisons of the average throughput incurred under the five relay selection policies. The parameter values used for figures figures are described in Table \ref{tab:parameters_fig_2_to_5}.
}
\label{fig:4}
\end{figure}

\begin{figure}[!t]
\centering
\subfloat[]{\includegraphics[width=0.235\textwidth]{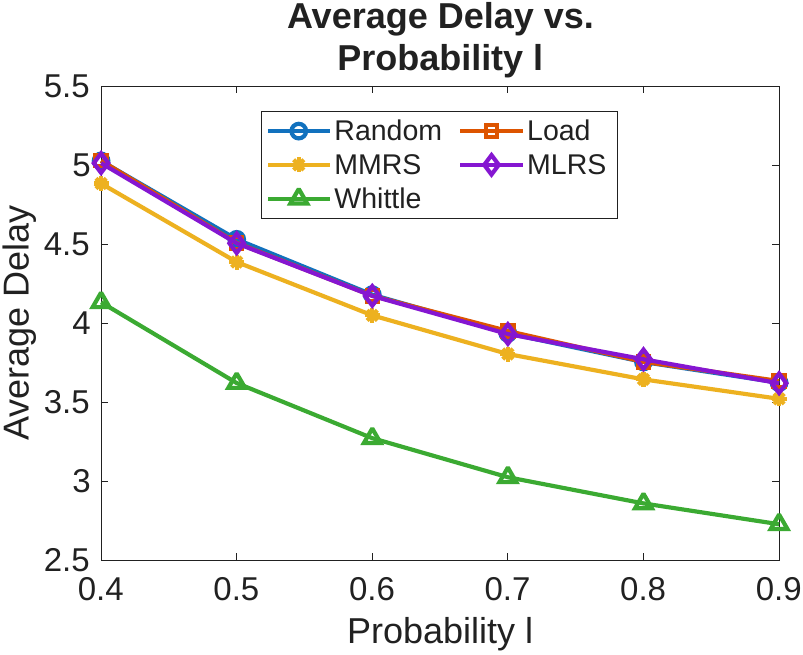}%
\label{fig:5a}}
\hfil
\subfloat[]{\includegraphics[width=0.24\textwidth]{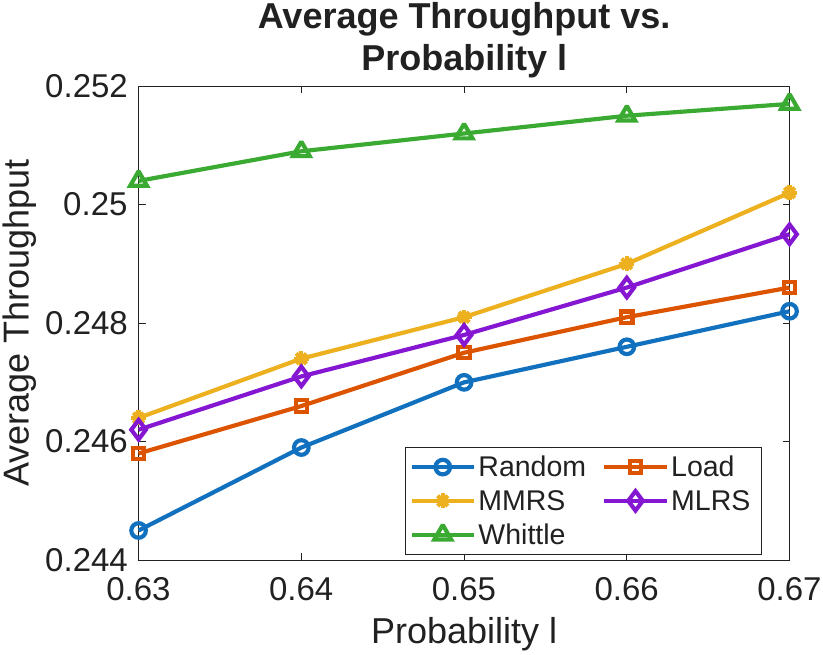}%
\label{fig:5b}}
\caption{The figures show comparisons of the average delay and average throughput incurred under the five relay selection policies. The parameter values used for figures figures are described in Table \ref{tab:parameters_fig_2_to_5}.
}
\label{fig:5}
\end{figure}

We assume that the packet queue at each relay is initially empty. 
Let $\textbf{f} = [f_1,  \ldots, f_M]$, $\textbf{l} = [l_1,  \ldots, l_M]$, and $\textbf{C} = [C_1,  \ldots, C_M]$, where $f_i, l_i$, and $C_i$ are defined as in Section \ref{system_model}. The parameter `buffer size' equals the maximum number of packets that can be stored in the queue of a relay. Table \ref{tab:parameters_fig_2_to_5} specifies the parameters required for all figs. \ref{fig:2} to \ref{fig:5}.

\begin{table}[]
 \begin{center}
 \caption{Parameters for Figures \ref{fig:2}-\ref{fig:5}}
 \label{tab:parameters_fig_2_to_5}
 \scalebox{0.7}{
\begin{tabular}{|clcclcclcc|}
\hline
\multicolumn{2}{|c|}{\multirow{2}{*}{Parameters}}                                                                                              & \multicolumn{8}{c|}{Value}                                                                                                                                                                                                                                                                                                                                                                                                                                                                            \\ \cline{3-10} 
\multicolumn{2}{|c|}{}                                                                                                                         & \multicolumn{1}{c|}{2(a)}                                              & \multicolumn{1}{c|}{2(b)}                                               & \multicolumn{1}{c|}{3(a)}                                      & \multicolumn{1}{c|}{3(b)}                                   & \multicolumn{1}{c|}{4(a)}                                     & \multicolumn{1}{c|}{4(b)}                               & \multicolumn{1}{c|}{5(a)}                              & 5(b)                             \\ \hline
\multicolumn{2}{|c|}{M}                                                                                                                        & \multicolumn{1}{c|}{5}                                                 & \multicolumn{1}{c|}{5 to 10}                                            & \multicolumn{1}{c|}{4 to 11}                                   & \multicolumn{1}{c|}{12}                                     & \multicolumn{1}{c|}{9 to 15}                                  & \multicolumn{1}{c|}{12}                                 & \multicolumn{1}{c|}{9}                                 & 4                                \\ \hline
\multicolumn{2}{|c|}{T}                                                                                                                        & \multicolumn{1}{c|}{30000}                                             & \multicolumn{1}{c|}{20000}                                              & \multicolumn{1}{c|}{20000}                                     & \multicolumn{1}{c|}{20000}                                  & \multicolumn{1}{c|}{20000}                                    & \multicolumn{1}{c|}{20000}                              & \multicolumn{1}{c|}{20000}                             & 20000                            \\ \hline
\multicolumn{2}{|c|}{Buffer Size}                                                                                                              & \multicolumn{1}{c|}{500}                                               & \multicolumn{1}{c|}{400}                                                & \multicolumn{1}{c|}{200}                                       & \multicolumn{1}{c|}{200}                                    & \multicolumn{1}{c|}{100}                                      & \multicolumn{1}{c|}{100}                                & \multicolumn{1}{c|}{100}                               & 50                               \\ \hline
\multicolumn{10}{l}{}                                                                                                                                                                                                                                                                                                                                                                                                                                                                                                                                                                                                                                \\ \hline
\multicolumn{1}{|c|}{\multirow{2}{*}{Figures}}                                                  & \multicolumn{9}{c|}{Parameters}                                                                                                                                                                                                                                                                                                                                                                                                                                                                                                                      \\ \cline{2-10} 
\multicolumn{1}{|c|}{}                                                                          & \multicolumn{3}{c|}{$\textbf{f}$}                                                                                                                                                                          & \multicolumn{3}{c|}{$\textbf{l}$}                                                                                                                                                                       & \multicolumn{3}{c|}{$\textbf{C}$}                                                                                                                              \\ \hline
\multicolumn{1}{|c|}{2(a)}                                                                      & \multicolumn{3}{l|}{{[}0.68,0.63,0.55,0.44,0.38{]}}                                                                                                                                             & \multicolumn{3}{l|}{{[}0.71,0.64,0.6,0.56,0.47{]}}                                                                                                                                           & \multicolumn{3}{l|}{{[}92,79,56,38,25{]}}                                                                                                           \\ \hline
\multicolumn{1}{|c|}{\begin{tabular}[c]{@{}c@{}}2(b) \\ (For M=5)\end{tabular}}                 & \multicolumn{3}{l|}{{[}0.62,0.59,0.56,0.53,0.5{]}}                                                                                                                                              & \multicolumn{3}{l|}{{[}0.63,0.6,0.57,0.54,0.51{]}}                                                                                                                                           & \multicolumn{3}{l|}{{[}91,90,89,88,87{]}}                                                                                                           \\ \hline
\multicolumn{1}{|c|}{\begin{tabular}[c]{@{}c@{}}3(a)\\ (For M=4)\end{tabular}}                  & \multicolumn{3}{l|}{{[}0.33,0.325,0.32,0.315{]}}                                                                                                                                                & \multicolumn{3}{l|}{{[}0.43,0.425,0.42,0.415{]}}                                                                                                                                             & \multicolumn{3}{l|}{{[}90,89.5,89,88.5{]}}                                                                                                          \\ \hline
\multicolumn{1}{|c|}{\multirow{4}{*}{3(b)}}                                                     & \multicolumn{3}{l|}{\multirow{4}{*}{\begin{tabular}[c]{@{}l@{}}$\textbf{f}$ is a vector of size $1 \times M$\\ with every element being f.\\ f varies from 0.1 to 0.8 \\ with an increment of 0.1.\end{tabular}}}   & \multicolumn{3}{l|}{\multirow{4}{*}{\begin{tabular}[c]{@{}l@{}}{[}0.889,0.886,0.883,0.88,\\ 0.877,0.874,0.871,0.868,\\ 0.865,0.862,0.859,0.856{]}\end{tabular}}}                             & \multicolumn{3}{l|}{\multirow{4}{*}{\begin{tabular}[c]{@{}l@{}}{[}90,89.9,89.8,89.7,\\ 89.6,89.5,89.4,89.3,\\ 89.2,89.1,89,88.9{]}\end{tabular}}}   \\
\multicolumn{1}{|c|}{}                                                                          & \multicolumn{3}{l|}{}                                                                                                                                                                           & \multicolumn{3}{l|}{}                                                                                                                                                                        & \multicolumn{3}{l|}{}                                                                                                                               \\
\multicolumn{1}{|c|}{}                                                                          & \multicolumn{3}{l|}{}                                                                                                                                                                           & \multicolumn{3}{l|}{}                                                                                                                                                                        & \multicolumn{3}{l|}{}                                                                                                                               \\
\multicolumn{1}{|c|}{}                                                                          & \multicolumn{3}{l|}{}                                                                                                                                                                           & \multicolumn{3}{l|}{}                                                                                                                                                                        & \multicolumn{3}{l|}{}                                                                                                                               \\ \hline
\multicolumn{1}{|c|}{\multirow{3}{*}{\begin{tabular}[c]{@{}c@{}}4(a)\\ (For M=9)\end{tabular}}} & \multicolumn{3}{l|}{\multirow{3}{*}{\begin{tabular}[c]{@{}l@{}}{[}0.795,0.792,0.789,0.786,\\ 0.777,0.783,0.781,0.779,\\ 0.778{]}\end{tabular}}}                                                 & \multicolumn{3}{l|}{\multirow{3}{*}{\begin{tabular}[c]{@{}l@{}}{[}0.8,0.797,0.794,0.791,\\ 0.788,0.786,0.784,0.782,\\ 0.783{]}\end{tabular}}}                                                & \multicolumn{3}{l|}{\multirow{3}{*}{\begin{tabular}[c]{@{}l@{}}{[}100,99.7,99.4,99.1,\\ 98.8,98.6,98.4,98.2,\\ 98.3{]}\end{tabular}}}               \\
\multicolumn{1}{|c|}{}                                                                          & \multicolumn{3}{l|}{}                                                                                                                                                                           & \multicolumn{3}{l|}{}                                                                                                                                                                        & \multicolumn{3}{l|}{}                                                                                                                               \\
\multicolumn{1}{|c|}{}                                                                          & \multicolumn{3}{l|}{}                                                                                                                                                                           & \multicolumn{3}{l|}{}                                                                                                                                                                        & \multicolumn{3}{l|}{}                                                                                                                               \\ \hline
\multicolumn{1}{|c|}{\multirow{4}{*}{4(b)}}                                                     & \multicolumn{3}{l|}{\multirow{4}{*}{\begin{tabular}[c]{@{}l@{}}$\textbf{f}$ is a vector of size $1 \times M$\\ with every element being f.\\ f varies from 0.20 to 0.25\\ with an increment of 0.01.\end{tabular}}} & \multicolumn{3}{l|}{\multirow{4}{*}{\begin{tabular}[c]{@{}l@{}}{[}0.65,0.647,0.644,0.641,\\ 0.638,0.635,0.633,0.631,\\ 0.629,0.627,0.625,0.623{]}\end{tabular}}}                             & \multicolumn{3}{l|}{\multirow{4}{*}{\begin{tabular}[c]{@{}l@{}}{[}60,59.7,59.4,59.1,\\ 58.8,58.5,58.3,58.1,\\ 57.9,57.7,57.5,57.3{]}\end{tabular}}} \\
\multicolumn{1}{|c|}{}                                                                          & \multicolumn{3}{l|}{}                                                                                                                                                                           & \multicolumn{3}{l|}{}                                                                                                                                                                        & \multicolumn{3}{l|}{}                                                                                                                               \\
\multicolumn{1}{|c|}{}                                                                          & \multicolumn{3}{l|}{}                                                                                                                                                                           & \multicolumn{3}{l|}{}                                                                                                                                                                        & \multicolumn{3}{l|}{}                                                                                                                               \\
\multicolumn{1}{|c|}{}                                                                          & \multicolumn{3}{l|}{}                                                                                                                                                                           & \multicolumn{3}{l|}{}                                                                                                                                                                        & \multicolumn{3}{l|}{}                                                                                                                               \\ \hline
\multicolumn{1}{|c|}{\multirow{4}{*}{5(a)}}                                                     & \multicolumn{3}{l|}{\multirow{4}{*}{\begin{tabular}[c]{@{}l@{}}{[}0.295,0.292,0.289,0.286,\\ 0.283,0.281,0.279,0.277,\\ 0.278{]}\end{tabular}}}                                                 & \multicolumn{3}{l|}{\multirow{4}{*}{\begin{tabular}[c]{@{}l@{}}$\textbf{l}$ is a vector of size $1 \times M$\\ with every element being l.\\ l varies from 0.4 to 0.9\\ with an increment of 0.1.\end{tabular}}} & \multicolumn{3}{l|}{\multirow{4}{*}{\begin{tabular}[c]{@{}l@{}}{[}100,99.7,99.4,99.1,\\ 98.8,98.6,98.4,98.2,\\ 98.3{]}\end{tabular}}}               \\
\multicolumn{1}{|c|}{}                                                                          & \multicolumn{3}{l|}{}                                                                                                                                                                           & \multicolumn{3}{l|}{}                                                                                                                                                                        & \multicolumn{3}{l|}{}                                                                                                                               \\
\multicolumn{1}{|c|}{}                                                                          & \multicolumn{3}{l|}{}                                                                                                                                                                           & \multicolumn{3}{l|}{}                                                                                                                                                                        & \multicolumn{3}{l|}{}                                                                                                                               \\
\multicolumn{1}{|c|}{}                                                                          & \multicolumn{3}{l|}{}                                                                                                                                                                           & \multicolumn{3}{l|}{}                                                                                                                                                                        & \multicolumn{3}{l|}{}                                                                                                                               \\ \hline
\multicolumn{1}{|c|}{5(b)}                                                                      & \multicolumn{3}{l|}{{[}0.25,0.248,0.25,0.248{]}}                                                                                                                                                & \multicolumn{3}{l|}{\begin{tabular}[c]{@{}l@{}}$\textbf{l}$ is a vector of size $1 \times M$\\ with every element being l.\\ l varies from 0.63 to 0.67\\ with an increment of 0.01.\end{tabular}}               & \multicolumn{3}{l|}{{[}60,59.8,59.9,59.8{]}}                                                                                                        \\ \hline
\end{tabular}}
\end{center}
\end{table}

 The performance metrics we use are as follows: long-run expected average packet holding cost at all the relays, average delay of packets, and average throughput. 
The delay of a packet is the number of time slots required for the packet to successfully reach the destination $D$ via one of the relays; it is measured starting from the time slot in which the packet reaches the head of the queue at the source $S$. 
The average of the delays of all the packets in a simulation run is the average delay. If $Q$ denotes the number of packets that have reached $D$ successfully in time $T$, then the average throughput is given by $Q/T$. 

Fig. \ref{fig:2a} shows the long-run expected average costs achieved under the five relay selection policies in time slots $20,001$ to $30,000$, whereas Fig. \ref{fig:2b} shows the long-run expected average costs versus the number of relays. Figs. \ref{fig:3a} and \ref{fig:3b} show the average delays under different relay selection policies versus the number of relays and versus the value of $f = f_1 = \ldots = f_M$, respectively.  Figs. \ref{fig:4a} and \ref{fig:4b} show the average throughput observed under different relay selection policies versus the number of relays and versus the value of $f$, respectively. Fig. \ref{fig:5} show the average throughput and average delay incurred under different relay selection policies versus different values of $l$. The results show that the proposed Whittle index-based policy consistently attains the best performance across all considered parameter values, and significantly outperforms all the other relay selection policies in terms of the average cost, average delay, as well as average throughput. 

\section{Conclusions And Future Work}\label{Section_conclusion}
We formulated the relay selection problem in wireless networks as a RMAB problem and proved that it is Whittle indexable. We proposed a method to compute the Whittle index of each relay in every time slot. In each time slot, our relay selection policy transmits a packet to the relay with the smallest Whittle index. Using simulations, we showed that the proposed policy outperforms relay selection policies proposed in prior work in terms of the  average cost, delay, as well as throughput. The results presented in this paper can be generalized as future work to encompass a scenario where multiple sources need to transmit files to some destinations, and the problem is to assign a relay from a set of candidate relays to each source. A primary insight in the case of multiple source–destination pairs is to prove that the formulated RMAB problem is Whittle indexable. The challenges we need to address to solve this problem are as follows: (i) \emph{Resource contention}: E.g., when it would be beneficial for two or more sources to use the same relay in a time slot, we need to design a mechanism that effectively resolves such contention and assigns a different relay to each of the sources. (ii) \emph{Inter-source interference mitigation}: The proposed scheme must incorporate mechanisms to mitigate the interference among the transmissions of different sources and different relays in a time slot. (iii) \emph{Fairness and efficiency tradeoff}: Overall throughput improves while serving single source-destination pair but it may also increase the delay for others. Also, centralized as well as distributed control policies can be designed to solve the above problem as part of future research.

\bibliographystyle{IEEEtran}

\begin{thebibliography}{10}
\providecommand{\url}[1]{#1}
\csname url@samestyle\endcsname
\providecommand{\newblock}{\relax}
\providecommand{\bibinfo}[2]{#2}
\providecommand{\BIBentrySTDinterwordspacing}{\spaceskip=0pt\relax}
\providecommand{\BIBentryALTinterwordstretchfactor}{4}
\providecommand{\BIBentryALTinterwordspacing}{\spaceskip=\fontdimen2\font plus
\BIBentryALTinterwordstretchfactor\fontdimen3\font minus \fontdimen4\font\relax}
\providecommand{\BIBforeignlanguage}[2]{{%
\expandafter\ifx\csname l@#1\endcsname\relax
\typeout{** WARNING: IEEEtran.bst: No hyphenation pattern has been}%
\typeout{** loaded for the language `#1'. Using the pattern for}%
\typeout{** the default language instead.}%
\else
\language=\csname l@#1\endcsname
\fi
#2}}
\providecommand{\BIBdecl}{\relax}
\BIBdecl

\bibitem{laneman2004cooperative} Laneman, J., Tse, D. \& Wornell, G. Cooperative diversity in wireless networks: Efficient protocols and outage behavior, {\em IEEE Trans. Inf. Theory}, vol. 50, pp. 3062-3080, Dec. 2004.

\bibitem{nam2008relay} Nam, S., Vu, M. \& Tarokh, V. Relay selection methods for wireless cooperative communications, {\em Proc. 42nd Annu. Conf. Inf. Sci. Syst.}, pp. 859-864, Mar. 2008.

\bibitem{kim2023joint} Kim, D., Castellanos, M. \& Heath, R. Joint relay selection and beam management based on deep reinforcement learning for millimeter wave vehicular communication, {\em IEEE Trans. Veh. Technol.}, vol. 72, pp. 13067-13080, May 2023.

\bibitem{ju2023deep} Ju, Y. \& Others Deep reinforcement learning based joint beam allocation and relay selection in mmWave vehicular networks, {\em IEEE Trans. Commun.}, vol. 71, no. 4, pp. 1997-2012, Jan. 2023.

\bibitem{ona2020relay} Onalan, A., Salik, E. \& Coleri, S. Relay Selection, Scheduling, and Power Control in Wireless-Powered Cooperative Communication Networks, {\em IEEE Trans. Wireless Commun.}. vol. 19, no. 11, pp. 7181-7195, Jul. 2020.

\bibitem{sheng2014energy} Sheng, Z. \& Others Energy-efficient relay selection for cooperative relaying in wireless multimedia networks, {\em IEEE Trans. Veh. Technol.}, vol 64, no. 3, pp. 1156-1170, May 2014.

\bibitem{zhang2020cooperative} Zhang, J., Tang, J. \& Wang, F. Cooperative relay selection for load balancing with mobility in hierarchical WSNs: A multi-armed bandit approach, {\em IEEE Access}, vol. 8, pp. 18110-18122, Jan. 2020.

\bibitem{li2023joint}Li, J. \& Others Joint optimization of relay selection and transmission scheduling for UAV-aided mmWave vehicular networks, {\em IEEE Trans. Veh. Technol.}, vol. 72, no. 5, pp. 6322-6334, Jan. 2023

\bibitem{gao2023optimal}Gao, H., Gu, X., Li, S., Di, Y. \& Vu, V. Optimal Relay Selection Method Based on Quantum Black Hole Algorithm, {\em Proc. 2nd Int. Conf. Computing Commun. Perception Quantum Tech. (CCPQT)}, pp. 1-5, 2023.

\bibitem{kequin2010indexability}Liu, K. \& Zhao, Q. Indexability of Restless Bandit Problems and Optimality of Whittle Index for Dynamic Multichannel Access, {\em IEEE Trans. Inf. Theory}, vol. 56, no. 11, pp. 5547-5567, Nov. 2010.

\bibitem{tripathi2019whittle}Tripathi, V. \& Modiano, E. A Whittle index approach to minimizing functions of age of information, {\em 57th Annu. Allerton Conf. Commun. Control Computing}, Monticello, IL, USA, pp. 1160-1167, 2019.

\bibitem{wang2019whittle}Wang, J., Ren, X., Mo, Y. \& Shi, L. Whittle index policy for dynamic multichannel allocation in remote state estimation, {\em IEEE Trans. Autom. Control}, vol. 65, no. 2, pp. 591-603, Feb. 2019.

\bibitem{papadimitriou1994complexity}Papadimitriou, C. \& Tsitsiklis, J. The complexity of optimal queueing network control, {\em Proc. IEEE 9th Annu. Conf. Struct. Complexity Theory}, Amsterdam, Netherlands, pp. 318-322, 1994.

\bibitem{whittle1988restless}Whittle, P. Restless bandits: Activity allocation in a changing world, {\em J. Appl. Probability}, vol. 25, pp. 287-298, 1988.

\bibitem{kaza2018restless}Kaza, K., Mehta, V., Meshram, R. \& Merchant, S. Restless bandits with cumulative feedback: Applications in wireless networks, {\em Proc. IEEE Wireless Commun. Netw. Conf. (WCNC)}, Barcelona, Spain, pp. 1-6, 2018.

\bibitem{hordijk1983average}Hordijk, A. \& Duyn Schouten, F. Average optimal policies in Markov decision drift processes with applications to a queueing and a replacement model, {\em Adv. Appl. Probability}, vol. 15, no. 2, pp. 274-303, 1983.

\bibitem{borkar2002convex}Borkar, V. Convex analytic methods in Markov decision processes, {\em Handbook Of Markov Decision Processes: Methods And Applications},Boston, MA, USA, Springer, 2002.

\bibitem{asmussen2003applied}Asmussen, S. Applied Probability and Queues, Springer-Verlag, 2003.

\bibitem{vsbspatta}Borkar, V. \& Pattathil, S. Whittle indexability in egalitarian processor sharing systems, {\em Ann. Operations Res.}, vol. 317, no. 2, pp. 417-437, 2022.

\bibitem{singh2022user}Singh, S., Borkar, V. \& Kasbekar, G. User Association in Dense mmWave Networks as Restless Bandits, {\em IEEE Trans. Veh. Technol.}, vol. 71, no. 7, pp. 7919-7929, Jul. 2022.

\bibitem{shen2021ber}Shen, M., Huang, Z., Lei, X. \& Fan, L. BER analysis of NOMA with max-min relay selection, {\em China Commun.}, vol. 18, no. 7, pp. 172-182, Jul. 2021.

\bibitem{el2020buffer}El-Rajab, M., Abou-Rjeily, C. \& Kfouri, R. Buffer-aided relaying: A survey on relay selection policies, {\em IET Commun.}, vol. 14, no. 21, pp. 3715-3734, Dec. 2020.

\end{thebibliography}

\end{document}